\documentclass[print,trackchanges]{ati}	
\usepackage{amsmath} 
\usepackage{amssymb} 
\usepackage{float}
\usepackage{xcolor}
\usepackage{ulem}
\usepackage{booktabs}  
\usepackage{multirow}   
\usepackage{array}      
\usepackage{booktabs}
\usepackage{graphicx,times}
\usepackage{lmodern}
\usepackage[sort&compress, numbers,super]{natbib} 
\bibpunct[,]{[}{]}{,}{s}{}{,}			
\usepackage{url}
\usepackage[colorlinks=true,breaklinks=true, linkcolor=red,urlcolor=magenta,citecolor=blue,anchorcolor=green]{hyperref}
\usepackage{caption}
\usepackage{rotating}
\usepackage{CJKutf8}	
\usepackage{fancyhdr}
\renewcommand{\citet}[1]{\citeauthor{#1}(\citeyear{#1})\cite{#1}}	

\begin{document}

   \title{
    The Analysis of the Influence of Coordinate Error of Observation Station On the Construction Accuracy of Pulsar Time
}


   \author{Zurong Zhou\correspondingAuthor{}
      \inst{1,2,3}
      \ORCID{0009-0005-8224-0677}		
   \and Chengshi Zhao\correspondingAuthor{}
     \inst{1,2}
   \and Yuping Gao
     \inst{1,2,3}
   \and Jianping Yuan
     \inst{3,4}
     \ORCID{0000-0002-5381-6498}
   \and Wei Han
     \inst{3,4}
   \and Shougang Zhang
     \inst{1,2,3}
   \and Yue Hu
     \inst{1,2,3}
   \and Shijun Dang
     \inst{5}
     \ORCID{0000-0002-2060-5539}
   \and Na Wang
     \inst{3,4}
     \ORCID{0000-0003-2991-7421}
   \and Jingbo Wang
     \inst{6}
   \and Minglei Tong
     \inst{1,2,3}
   \and De Wu
     \inst{1,2,3}
   }
\correspondent{Chengshi Zhao}	
\correspondentEmail{zhaocs@ntsc.ac.cn;zhouzurong@ntsc.ac.cn}
\institute{ National Time Service Center, Chinese Academy of Sciences, Xi'an 710600, China;
          \and
              Key Laboratory of Time Reference and Applications,Chinese Academy of Sciences, Xi'an 710600, China;
          \and
              University of Chinese Academy of Sciences,BeiJing 100049, People's Republic of China;
          \and
              Xinjiang Astronomical Observatory, Chinese Academy of Sciences, Urumqi 830011, People's Republic of China;
          \and
              Guizhou Radio Astronomical Observatory, Guizhou University, Guiyang 550001, People's Republic of China;
          \and
              Lishui University,Lishui 323000, People's Republic of China;
   }
   \date{Received:~December 23, 2025;   Accepted:~April 22, 2026;  Published Online:~xxxx xx, 2026; 
   \DOI{ati2024999} }			
   \citeinfo {Zhou, Z.-R. et al.}\volume{1}\issue{5} \pages{560--573}	
   \StartPage{560} 			
   \MonthIssue{December}		
   \copyrights {2025}     		
   \abstract{Errors in observatory coordinates directly impact the precision of pulsar time-scale construction. Using the pulsar timing software TEMPO2, this study simulates various station position errors within the three-dimensional terrestrial reference frame for three different types of millisecond pulsars, over periods of 13 days and 5 years, and analyzes their effects on pulsar timing results.The findings demonstrate that, for both 13‑day and 5‑year observation spans, station coordinate errors substantially reduce the accuracy of pulsar timescale construction when the zenith angle exhibits long‑term variations. This effect is independent of pulsar type and the daily observable time of the station antenna for the pulsar. A linear relationship is found between station coordinate errors and the Root-Mean-Square (RMS) of pulsar timing residuals, with fitted linear coefficients ranging from $1.36\times{10}^{-11}$ to $1.61\times{10}^{-9}$ for the three pulsars. The Roemer delay error caused by coordinate inaccuracies is notably larger than other delay and correction terms. Errors along the $x$- and $y$-axes have comparable influences on timing precision, whereas errors along the $z$-axis have a relatively smaller effect. Kendall correlation analysis between station error-induced Roemer delay and RMS yields a correlation coefficient $r=1.67\%$ and $p=100\%$ in all cases, indicating that, at current timing precision levels, coordinate errors primarily affect the Roemer delay term and thus the pulse arrival times, which is highly consistent with theoretical models.While these findings offer valuable insights into the key factors influencing pulsar timescale accuracy and related applications, they may not hold under conditions of a constant zenith angle or limited elevation angles, such as those at FAST.
  }
   \authorrunning{ASTRONOMICAL TECHNIQUES \& INSTRUMENTS }   
   \titlerunning{Zhou Z.-R. et al.: ~Prepare a LaTeX Manuscript for ATI }  
   \maketitle
   \setcounter{page}{\Page}	
%
%
\section{Introduction}
Pulsars are a class of neutron stars characterized by extremely stable rotation, with millisecond pulsars in particular often regarded as the most stable "astronomical clocks" in nature\citep{Taylor1991}. Leveraging the highly stable rotational properties of millisecond pulsars, it is possible to establish an ideal time scale independent of atomic time, known as Pulsar Time (PT). With continuous advancements in science and technology, the observational capabilities of radio telescopes have significantly improved, leading to the discovery of more millisecond pulsars and greatly promoting the application of Pulsar Time in fields such as the establishment of pulsar time standards\citep{Hobbs2020} and pulsar navigation\citep{Zhao2011,Shuai2019}. Pulsar timing\citep{Hobbs2009} is a technical approach that involves regular monitoring of the Time of Arrival (TOA) of pulses through observatories (telescopes or spacecraft)\citep{Yang2008} to develop physical models\citep{Zhou2007}, enabling the construction and application of Pulsar Time. Conversely, a stable and reliable Pulsar Time can provide long-term constraints for pulsar timing models and support broader applications in spacetime reference systems\citep{Han2023}. The accuracy of pulsar timing is influenced by various error sources, including time comparison system errors, inaccuracies in Earth ephemerides, uncertainties in solar system body masses, gravitational wave effects, measurement noise arising from radiometer noise and pulse phase jitter, timing red noise caused by intrinsic pulsar rotational instability, red noise induced by dispersion measure variations in the interstellar medium, errors in timing models and algorithms, as well as noise from observational systems\citep{Ding2017}.

With the rapid advancement of space geodetic techniques such as Very Long Baseline Interferometry (VLBI) and the Global Navigation Satellite System (GNSS), errors in ground station coordinates within the International Terrestrial Reference Frame (ITRF) have been well controlled. Although these errors are not a dominant factor, their impact on pulsar timing has not been completely eliminated. Due to factors such as satellite orbit and signal propagation delays, biases in receiving equipment, site deformations caused by Earth tides, atmospheric and hydrological loading \citep{Zhu2020}, as well as extreme conditions including imperfect orbital dynamic models, measurement system errors, spatiotemporal reference frame deviations, and thermal deformation, attitude, and control errors of the spacecraft platform itself \citep{Liu2025}, discrepancies ranging from several millimeters \citep{Guillory2023,Ma2023} to tens of kilometers \citep{Tian2011} exist between the theoretical station position used for calculating TOA and the true position required by the physics of signal propagation. These discrepancies can subsequently affect applications such as pulsar time across different timescales. Pulsar navigation technology holds significant value for both military and civilian applications \citep{Sheikh2007,Emadzadeh2011}. From an astrometric perspective, \citeauthor{Li2018} have also emphasized the necessity of considering station position errors in pulsar navigation research \citep{Li2018}. Recent findings by D. Kaur and G. Hobbs demonstrate that a small radio antenna with a diameter of 4 meters, operating around 700 MHz, can not only detect more pulsars but also utilize pulsar navigation technology to determine its own position with an accuracy of about 10 kilometers \citep{Kaur2025}.

To investigate the application of pulsar time, this paper focuses on the impact of station coordinate errors on the precision of pulsar time-scale construction. A mathematical relationship between the positional errors of the station reference point and various indicators of pulsar time-scale precision is first established at the theoretical level. Subsequently, data simulation methods are employed for verification and analysis. The study aims to provide a theoretical foundation and methodological guidance for the future inversion of station coordinate errors based on pulsar time-scale precision.

\section{Theoretical Analysis of the Influence of Station Coordinate Errors on Pulsar Time}

In pulsar timing observations, the clock of the ground station or satellite is typically selected as the time reference. Its time must be corrected to an international standard time system, such as International Atomic Time (TAI) or Terrestrial Time (TT), to ensure that the timing observations are referenced to the highest-precision time system available today. To avoid the effects of Earth kinematics and dynamics, the processing and analysis of pulsar observation events in near-Earth space require the selection of an ideal inertial reference system. The Solar System Barycenter (SSB) is generally chosen as this reference frame. Therefore, the time of arrival measured at the station must be transformed to the TOA at the SSB, with the time also converted to Barycentric Coordinate Time (TCB). Taking ground-based radio telescope timing observations as an example, after receiving pulsar signals, the telescope accounts for a series of physical, geometric, and relativistic effects, including flat-space geometric delays, gravitational delays, time-scale effects, and propagation delays through various media. The transformation from the station TOA ($t_{obs}$) to the TOA at the SSB ($t_{SSB}$) requires correcting for these effects, as described by the following mathematical model \citep{Tong2017}:

\begin{equation}
\begin{aligned}
t_{\mathrm{SSB}} = t_{\mathrm{obs}} 
&+ \frac{\vec{n}\cdot\vec{r}}{c} - \frac{1}{2c R_{0}} \left[ (\vec{r})^{2} - (\vec{n}\cdot\vec{r})^{2} \right] \\
&-\frac{1}{c}(\vec{n}\cdot\vec{v})\Delta t + \frac{1}{c R_{0}} \left[ (\vec{r}\cdot\vec{v}) - (\vec{n}\cdot\vec{r})(\vec{n}\cdot\vec{v}) \right] \Delta t \\
&+ \Delta t_{c} + \frac{2}{c^{3}} \sum_{k=S,E\cdot\cdot\cdot} \mu_{k} \ln \left( r_{k} + \vec{n}\cdot\vec{r}_{k} \right) \\
&- \frac{4\mu_{S}^{2}}{c^{5}r_{S}\tan\psi\sin\psi} - \frac{D}{f^{2}} - \Delta_{B}
\end{aligned}
\label{eq1}
\end{equation}

In the equation, $\vec{n}$ is the unit direction vector of the pulsar in the SSB frame, $\vec{r}$ is the position vector of the observational station (telescope) relative to the SSB at the time of observation, $\vec{v}$ is the velocity vector of the pulsar relative to the SSB, $c$ is the speed of light, and$\Delta t$ is the time difference between the observation epoch and the reference epoch, i.e.,$\Delta t = t - t_{0}$. $R_{0}$ is the distance from the pulsar to the SSB at the reference epoch, $G$ is the gravitational constant, and $\Delta t_{c}$ represents the clock correction term converting the telescope's time to the TCB scale, including atomic clock offsets and various Einstein delays (e.g., transformations between TAI, TT, and TCB). $\mu_{k}$ is the gravitational constant of the $k$-th solar system body, $\vec r_{k}$ is the position vector of the $k$-th solar system body relative to the radio telescope, and $r_{k}$ is its magnitude. $\psi$ is the angular separation between the Sun and the pulsar as seen from the telescope, $D$ is the dispersion measure of the pulsar, and $f$ is the observing frequency at the SSB. $\Delta B$ denotes the signal propagation delay within a binary system, mainly including coordinate transformation to the pulsar frame, vacuum propagation delay due to orbital motion, propagation delay caused by the signal passing through the companion's gravitational field, and relativistic corrections to the time coordinate\citep{Zhou2007,Tong2017}.The terms 2 to 5 on the right-hand side of Equation \ref{eq1} represent the geometric vacuum delay, parallax, delay due to the pulsar's radial velocity, and the first-order Roemer delay, respectively. Terms 7 to 9 correspond to various gravitational delays in the solar system, second-order solar gravitational delay, and dispersion delay. For space-based X-ray pulsar data processing, since X-rays are electromagnetic waves with short wavelengths and high energy, their dispersion delay can be neglected.

\begin{figure}
    \begin{minipage}[t]{0.999\linewidth}  
    \centering
    \includegraphics[scale=0.42]{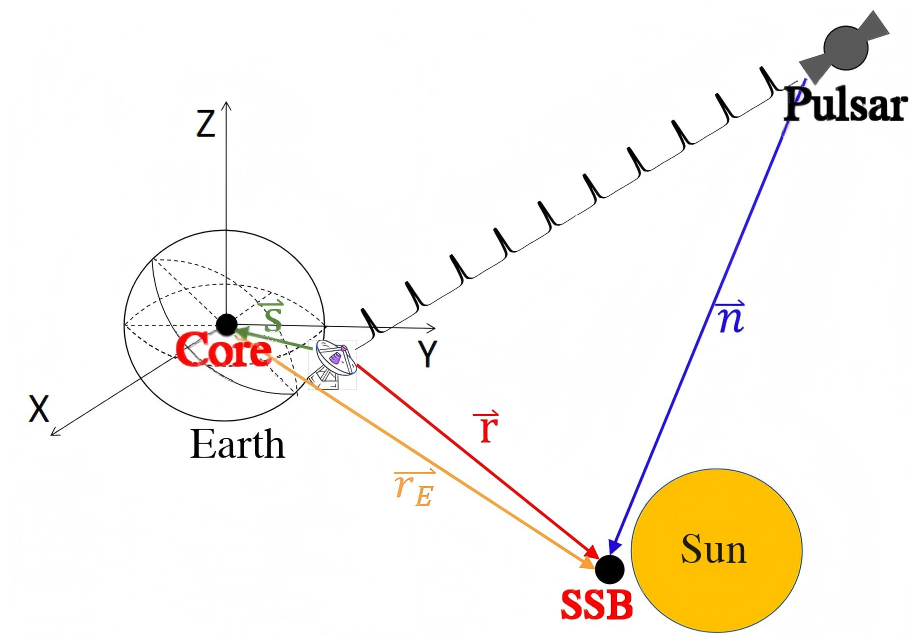} 
    \caption{A diagram of pulsar signal propagation to the observatory of earth}
    \label{fig1}
    \end{minipage}
\end{figure}

By analyzing $t_{SSB}$, a pulsar timing model (i.e., a pulsar clock model) can be established, enabling precise prediction of pulse arrival times\citep{Edwards2006}. The timing residual ($R_{res}$) is defined as the difference between the predicted and observed pulse arrival times at the solar system barycenter, expressed as PT$-$TT. Typically, model parameters are optimized through fitting to better align the timing model with the observational data. Ideally, if the timing model perfectly describes all factors influencing the TOA, the timing residuals follow a Gaussian distribution with zero mean, and their standard deviation matches the measurement error of the arrival times. If the timing model inadequately captures the characteristics of the TOA, the residuals will deviate from a Gaussian distribution and may exhibit specific systematic features. Therefore, timing residuals serve as a crucial indicator for assessing the accuracy of the pulsar timing model, and their RMS value is commonly used to quantify the measurement precision and stability of the pulsar time. In the pulsar timing model, the relationship between observatory coordinates and the model is not entirely linear. Specifically, the influence of the Shapiro delay, Einstein delay, higher-order terms of the Roemer delay, and all their partial derivatives on station coordinates can be neglected. The magnitude of the zeroth-order Roemer delay term $\frac{\vec{n}\cdot \vec{r}}{c}$ is significantly larger than that of other terms. Consequently, the primary Roemer delay term considered in Equation \ref{eq1} can be approximated as $\Delta R_{sun}=-\frac{\vec{n}\cdot \vec{r}}{c}$. This term depends on the pulsar's direction vector and the station's position vector $\vec{r}=\vec{s}+\vec{r_{E}}$ relative to the SSB, where $\vec{s}$ denotes the station's position vector relative to the geocenter, and $\vec{r}_E$ represents the geocenter's position vector relative to the SSB, as illustrated in Figure \ref{fig1}. This paper treats Earth's annual motion around the SSB as an elliptical orbit with an eccentricity $e = 0.0167$ and an orbital period of one year.Evidently,$\frac{\vec{n}\cdot \vec{s}}{c}$ exhibits diurnal variation, while$\frac{\vec{n}\cdot \vec{r_{E}}}{c}$demonstrates annual variation. The magnitude of $\frac{\vec{n}\cdot \vec{s}}{c}$is approximately five orders of magnitude smaller than that of $\frac{\vec{n}\cdot \vec{r_{E}}}{c}$. Therefore, the Roemer delay error induced by station position errors can be approximately expressed as:

\begin{equation}
\Delta R_{sun}= -\frac{\vec{n}\cdot \vec{s}}{c}-\frac{\vec{n}\cdot \vec{r_{E}}}{c}
\label{eq2}
\end{equation}

Therefore, the relationship between pulsar timing residuals and station position coordinates can be approximated as$ R_{\text{res}} \approx d\Delta R_{\text{sun}}$, where $ R_{\text{res}}$ represents the residuals before fitting, and $d\Delta R_{\text{sun}}$ denotes the first-order differential of the Roemer delay term. In the International Celestial Reference System (ICRS), timing observations of a single pulsar allow the determination of a reference position along the pulsar’s line of sight. For timing observations of two pulsars, the position is confined to the intersection line of two planes. When timing observations involve three or more pulsars, the position can be precisely determined in three-dimensional space. The above relationship can be concisely expressed as follows:

\begin{equation}
R_{res}\approx d \Delta R_{sun}=-\frac{\vec{n}\cdot \vec{\Delta{r}}}{c}
\label{eq3}
\end{equation}

Here, $\Delta\vec{r}$ represents the error in the observer’s position. In the International Terrestrial Reference System (ITRS), the relationship between pulsar timing residuals and station coordinates can be derived based on Equation \ref{eq3} and the transformation between ITRS and the Geocentric Celestial Reference System (GCRS)\citep{Han2019}:

\begin{equation}
\begin{aligned}
R_{\mathrm{res}} = &\left(\frac{\partial\Delta R_{\mathrm{sun}}}{\partial X_{\mathrm{ITRS}}}\right)\Delta X_{\mathrm{ITRS}} 
                   + \left(\frac{\partial\Delta R_{\mathrm{sun}}}{\partial Y_{\mathrm{ITRS}}}\right)\Delta Y_{\mathrm{ITRS}} \\
                  &+ \left(\frac{\partial\Delta R_{\mathrm{sun}}}{\partial Z_{\mathrm{ITRS}}}\right)\Delta Z_{\mathrm{ITRS}}
\end{aligned}
\label{eq4}
\end{equation}

Here, $X_{ITRS}$,$ Y_{ITRS}$, and$ Z_{ITRS}$ denote the spatial coordinates of the observing station in the International Terrestrial Reference System (ITRS). It should be noted that, due to the influence of Earth’s rotation, $\partial \Delta R_{\text{sun}} / \partial X_{ITRS} $ and $\partial \Delta R_{\text{sun}} / \partial Y_{ITRS}$ exhibit temporal variations, while $ \partial \Delta R_{\text{sun}} / \partial Z_{ITRS}$ (approximately equal to $\sin\delta/c$, where $\delta$ is the declination of the pulsar and $c$ is the speed of light) remains essentially constant.

Due to differences in station location, the Earth’s gravitational potential and rotation velocity vary, resulting in corrections from TT to TCB that are not identical at the same epoch. Within the framework of the ICRS, the conversion from TT to TCB is given as follows\citep{Luzum2011}:

%

\begin{equation}
\begin{split}
\mathrm{TCB}=\mathrm{TT}+c^{-2}\Biggl\{\int_{t_{0}}^{t}\Bigl[v_{e}^{2}/2+U_{\mathrm{ext}}(\mathbf{x}_{e})\Bigr]\,\mathrm{d}t & \\
+\mathbf{v}_{e}\cdot(\mathbf{x}-\mathbf{x}_{e})\Biggr\} & \\
+O\left(c^{-4}\right)+L_{G}\times(M J D-43\,144.0)\times86400\;s&
\label{eq5}
\end{split}
\end{equation}

where \(\mathbf{x}\) and \(\mathbf{v}_e\) are the position and velocity of the Earth’s center relative to the barycenter of the solar system, \(U_{ext}\) denotes the Newtonian gravitational potential at the geocenter due to all solar system bodies except the Earth itself, the \(O(c^{-4})\) term is smaller than \(10^{-16}\) in rate and can be neglected, \(L_G = 6.969290134\times10^{-10}\) is a defining constant, and MJD is the Modified Julian Date.

\section{Selection of Pulsars and Simulation Methods}

\begin{figure}[H]
\begin{minipage}[t]{0.999\linewidth}
\centering
\includegraphics[scale=0.40]{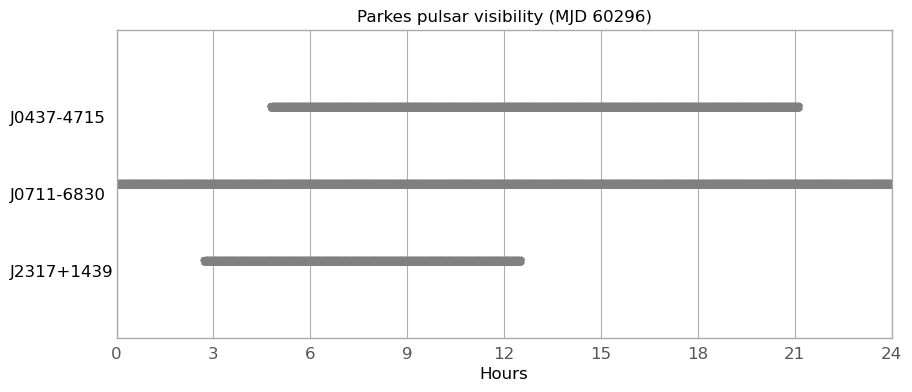}
\caption{Time distribution of three millisecond pulsars in the sky region of the Parkes 64-meter telescope.}
\label{fig2}
\end{minipage}
\end{figure}

Taking ground-based observatories as an example,this study selects three pulsars(PSRs J0437$-$4715, J0711$-$6830, and J2317$+$1439) from the data set of International Pulsar Timing Array (IPTA)\citep{Hobbs2010} with high-precision TOA data as subjects for simulation analysis. Their published timing residuals have been fitted using the TEMPO2 software\citep{Zic2023,Perera2019}, and their relevant basic parameters are detailed in Table~\ref{Table1}. These three pulsars are distributed across different regions of the northern and southern skies: J0711$-$6830 is an isolated millisecond pulsar, while J0437$-$4715 and J2317$+$1439 belong to distinct binary pulsar systems, exhibiting significant differences in orbital period and eccentricity. Among them, J0437$-$4715 is the brightest, whereas J2317$+$1439 is relatively faint.The IPTA mainly includes the Parkes Pulsar Timing Array (PPTA)\citep{Reardon2023}, the European Pulsar Timing Array (EPTA)\citep{EPTA2023}, and the Chinese Pulsar Timing Array (CPTA)\citep{Xu2023}. According to observational results from PPTA DR3, J0437$-$4715 and J0711$-$6830 demonstrate higher pulsar timing precision\citep{Zic2023}, while data from IPTA DR2 indicate that the timing precision of J2317$+$1439 lies between the other two\citep{Perera2019}. Based on the spatial distribution of these pulsars within the Galaxy, their observable durations in the scanning area of the Parkes 64-meter telescope on MJD 60296 were calculated to be approximately 10 hours, 17 hours, and 24 hours, respectively, as illustrated in Figure~\ref{fig2}.

\begin{table*}
\caption[]{The basic parameters of the three selected millisecond pulsars.}
\begin{tabular}{ccccccccccc}
\hline
 Pulsar name & RA & DEC & P & DM & Binary & Pb & Ecc & $S_{1400}$ &Residuals& References \\ 
 (J2000)&$rad$&$rad$&$(ms)$&$(cm^{-3}pc)$& &$(d)$& &$(mJy)$&$(\mu s)$ &for position \\ \hline
 J0437$-$4715&1.210&$-0.825$&5.8&2.64&T2 &5.7&$1.918\times{10}^{-5}$&150.2&0.36&\cite{Zic2023}\\
 J0711$-$6830&1.885&$-1.196$&5.5&18.41&$-$&$-$&$-$&2.6&0.91&\cite{Zic2023}\\
 J2317$+$1439&6.093&0.256 &3.4 &21.90&ELL1H &2.46&$5.4\times{10}^{-7}$&0.60&0.88&\cite{Perera2019}\\
\hline
\end{tabular}
\label{Table1}
\end{table*}

Currently, numerous pulsar timing models have been integrated into pulsar timing software packages, among which TEMPO2 is widely adopted\citep{Edwards2006}. Compared to earlier timing software, TEMPO2 employs the general relativistic framework recommended by the International Astronomical Union (IAU) in 1991 and 2000, based on the ICRS, TCB, and the latest models for precession, nutation, and polar motion\citep{Edwards2006}. In contrast to other timing software such as TEMPO\citep{Nice2015}and PSRTIME\citep{Doroshenko1990}, TEMPO2 incorporates improvements and updates in both theoretical foundations and algorithmic implementations\citep{Hobbs2006}, and it is now extensively used in data processing for pulsar timing and its applications.This study utilizes the TEMPO2 software to simulate and analyze the impact of errors in the station coordinate reference point on pulsar timing accuracy. In the terrestrial coordinate system, the site coordinates of the Parkes Observatory are $(-4554231.5, 2816759.1, -3454036.3)$ meters along the $x$-, $y$-, and $z$-axes, respectively, with high measurement precision. By adjusting the values of the three coordinate components individually as $\Delta x$,  $\Delta y$, and  $\Delta z$, the influence of station coordinate errors on timing residuals and various delay terms is simulated. Given that actual station coordinate variations are typically not confined to a single direction, this study also simulates the effect on pulsar timing accuracy when the station reference point shifts simultaneously in all three directions. Based on measurement tests of ground and space station coordinates, the maximum error in a single coordinate direction can reach up to 10 km. Therefore, this simulation primarily considers error magnitudes of 1 , 10 , 100 , and 1000 m in each of the three coordinate directions.

Typically, based on ephemeris parameter files, a set of TOA sequence data under the condition of no coordinate errors is first simulated for each pulsar. Such data sets can be used for algorithm testing. The TEMPO2 software defaults to using the Parkes Observatory as the observation site, with the observing frequency set to 1440 MHz. All data are regularly sampled and share the same uncertainty. The simulation in this study is divided into two categories: a short-term simulation with daily sampling over 13 days, and a long-term simulation with yearly sampling over 5 consecutive years. In the 13-day continuous sampling, each data set consists of 240 observations per day at random hour angles within a 24-hour period, covering the time range from MJD 60290 to MJD 60302. In the 5-year long-term sampling, each data set includes one observation per day at a random hour angle within 24 hours, spanning the period from MJD 58475 to MJD 60301.When the observation hour angle is randomly covered, this means that the zenith angles of the two types of observations for these three pulsars vary arbitrarily within a certain specific range. To prevent interference from other factors with station error, this experiment employed a controlled variable approach. The uncertainty of each arrival time was set to 0.1 ns, thereby achieving ideal fitting residuals for all three pulsars, with the Root Mean Square (RMS) of the residuals being 0 $\mu$s. This indicates that the pulsar timing precision was unaffected by any external factors under these conditions.After confirming the accuracy of the ephemeris and timing parameters for each pulsar, the timing parameters are further computed by incrementally adjusting the three-dimensional coordinates of the station. These parameters include the TOA, residual RMS, Roemer delay term, Shapiro delay term, Einstein delay term, and correction from TT to TCB. 

Owing to the elevation angle constraints of a radio telescope, the zenith angle for any practical observation must remain below the telescope’s maximum allowable zenith angle.Neither of the two types of simulations described above considered the effect of variations in the maximum observable zenith angle on pulsar timing. To ensure the completeness of the experiments and conclusions, we additionally performed a five‑year simulation to investigate the effect of station coordinate errors on the construction of pulsar timescales for the three pulsars, under the assumption of a constant zenith angle. Except for fixing the zenith angle, all other simulation settings were identical to those used in the second category. As the TEMPO2 software includes an internal atmospheric delay model capable of correcting the hydrostatic tropospheric delay at the level of 10 ns\citep{Edwards2006}, and additionally provides a 1.5 ns correction for the zenith wet delay\citep{Hobbs2006},both effects are therefore omitted from the simulations and discussions presented herein.

\section{Simulation Results}
Through multiple experiments and statistical analyses, this study obtained simulation results for three millisecond pulsars over both a continuous 13-day period and a continuous 5-year period. During the 13-day simulation, each pulsar generated 2640 TOAs, while 1832 TOAs were obtained during the 5-year simulation. As shown in the following figure of residual plots, when the zenith angle changes arbitrarily, the simulated TOAs do not exhibit systematic errors during periods when the pulsars are outside the telescope's observable sky region, nor do the results show significant differences attributable to the different types of millisecond pulsars. This indicates that using the TEMPO2 software to simulate pulsar TOAs produces results in which the number of TOAs and timing outcomes are independent of whether the pulsar is within the telescope's observable sky region or of the pulsar's type. The experimental results further demonstrate that when station coordinate errors are small, the pulsars' RMS values and Roemer delay errors are small ,and the error in the correction for the TT-to-TCB conversion is negligible and close to zero . As station coordinate errors increase, the pulsars' RMS values and Roemer delay errors increase significantly ,and the error in the correction for the TT-to-TCB conversion also appears to increase slowly. However, this conclusion does not appear to hold when the zenith angle remains constant.Specific data and analysis results are detailed in Tables 2–3 and Figures 3-12.

\subsection{Analysis of the Simulation Results for 13-Day Continuous Sampling Data}
Table~\ref{Table2} summarizes the statistical results of the impact of various station coordinate errors on three millisecond pulsars during 13 days of continuous timing observations, under the assumption that the zenith angle varies within a certain range. The first column lists, under different station coordinate error conditions, the obtained RMS values, the ranges of Roemer delay errors, pre-fit residual errors, and correction errors from TT to TCB, along with their corresponding units. The second column specifies the pulsar names. Columns three to seven present the specific numerical values of each error metric under five different levels of coordinate errors. As shown by the data, as the station coordinate errors increase, the RMS, the range of Roemer delay errors, and the range of correction errors from TT to TCB for all three pulsars exhibit a linear increasing trend. The specific variations of each parameter will be discussed in detail in subsequent sections.

\begin{table*}
\center
\caption[]{The statistical analysis of the influence of station error on the 13 days pulsar timing of different pulsars when zenith angle changes arbitrarily.}
\begin{tabular}{ccccccc}
\hline
\multirow{2}{*}{Categories}& Psr Name &\multicolumn{5}{c}{($\Delta x$,$\Delta y$,$\Delta z$)(m,m,m)}\\
\cmidrule(lr){3-7}
   & (J2000) &(0,0,0) &(1,1,1) & (10,10,10)& (100,100,100)& (1000,1000,1000)\\
\hline
\multirow{3}{*}{RMS}&J0437$-$4715&0&0.002&0.023&0.226&2.258\\
  \quad &J0711$-$6830&0 &0.001&0.012&0.121&1.214 \\
 ($\mu$s) &J2317$+$1439&0 &0.003&0.032&0.321&3.213\\
\multirow{3}{*}{$\Delta R_{roemer}$}&J0437$-$4715&0 &6.41&64.10&640.98&6409.81\\
  \quad &J0711$-$6830&0 &3.45&34.50&344.98&3449.83 \\
 (ns)&J2317$+$1439&0 &9.12 &91.22 &912.21 &9122.07\\
\multirow{3}{*}{$\Delta R_{pre-res}$}&J0437$-$4715&0 &6.76 &64.77 &641.25 &6410.06\\
  \quad &J0711$-$6830&0 &3.68 &34.69 &345.32 &3450.48\\
 (ns) &J2317$+$1439&0 &9.53&91.49&912.57&9122.08 \\
\multirow{3}{*}{$\Delta R_{TT to TCB}$}& J0437$-$4715&0&0&0.01 &0.10 &0.97\\
 \quad &J0711$-$6830&0&0 &0.01 &0.10 &0.97\\
 (ns) &J2317$+$1439&0&0 &0.01&0.10 &0.97\\
\hline
\end{tabular}
\label{Table2}
\end{table*}

\subsubsection{TOA and Timing Residuals}

\begin{figure}
\begin{minipage}[t]{0.999\linewidth}
\centering
\includegraphics[scale=0.20]{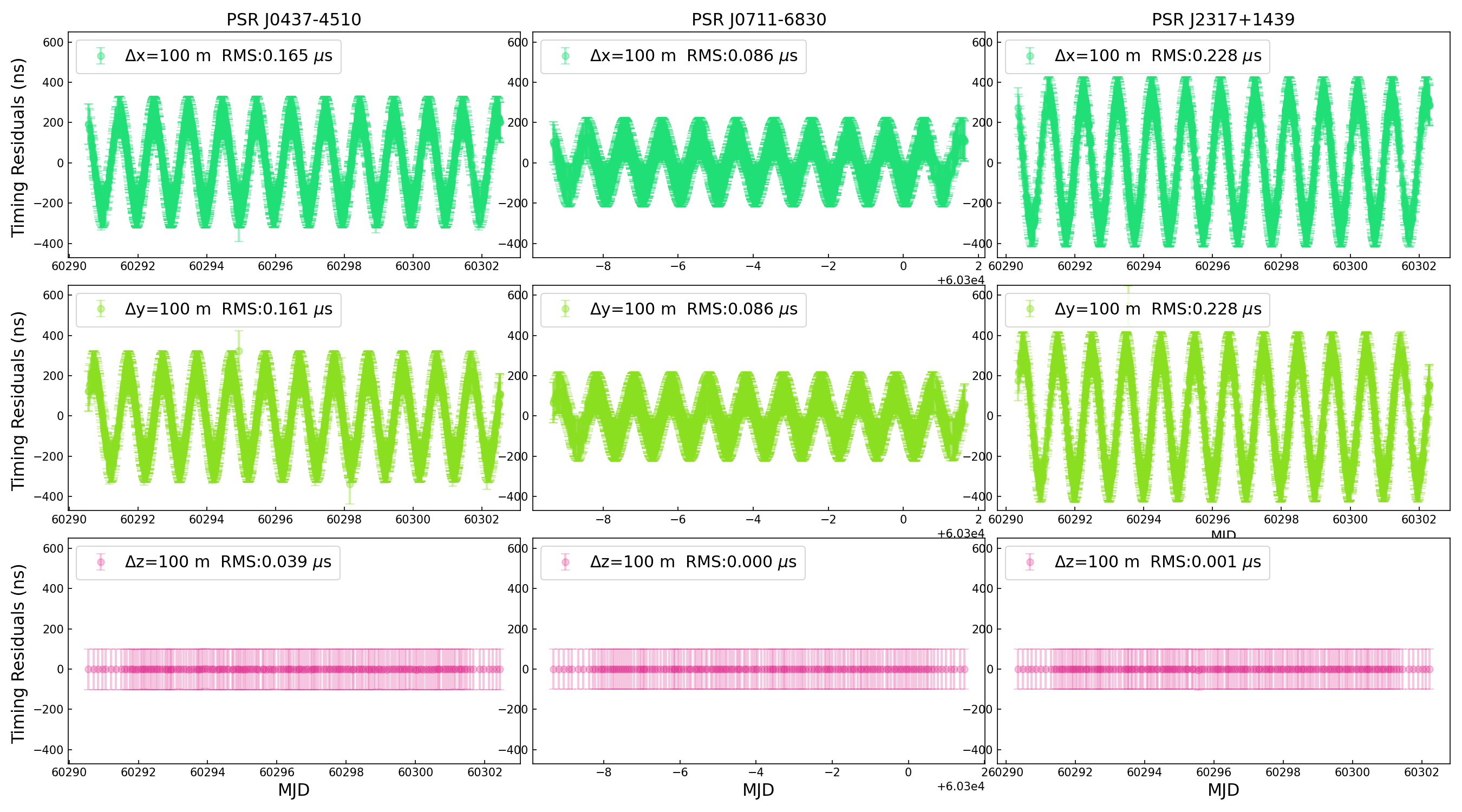}
\caption{The influence of station error on 13-day timing residual when zenith angle changes.}
\label{fig3}
\end{minipage}
\end{figure}

Based at the Parkes Observatory in Australia, this study analyzes the impact of station coordinate errors on the Time of Arrival (TOA) and its residuals for pulsars J0437$-$4715, J0711$-$6830, and J2317$+$1439 ,under the assumption that the zenith angle changes within a certain range. Assuming a coordinate error of 100 metres in each of the three spatial dimensions, Figure~\ref{fig3} illustrates the variations in TOA errors, residual distributions, and RMS values for each pulsar under three different error directions. As shown in Figure~\ref{fig3}, when the station coordinate error is 100 metres along in the $x$, $y$, and $z$ directions, the induced RMS values are comparable between the$x$- and $y$-axis directions and are significantly larger than those in the $z$-axis direction. Regarding the amplitude of the RMS value induced by coordinate errors along the $x\text{/}y$ axes, J2317$+$1439 exhibits significantly larger values than J0437$-$4715, while J0711$-$6830 shows the smallest amplitude. Specifically, the TOA error in the $x$-axis direction exhibits a cosine waveform,the $y$-axis direction shows a sinusoidal waveform, while the TOA error in the $z$-axis direction is uniformly distributed above and below the zero‑residual baseline.

\subsubsection{Roemer Delay}

\begin{figure}[H]
\begin{minipage}[t]{0.999\linewidth}
\centering
\includegraphics[scale=0.20]{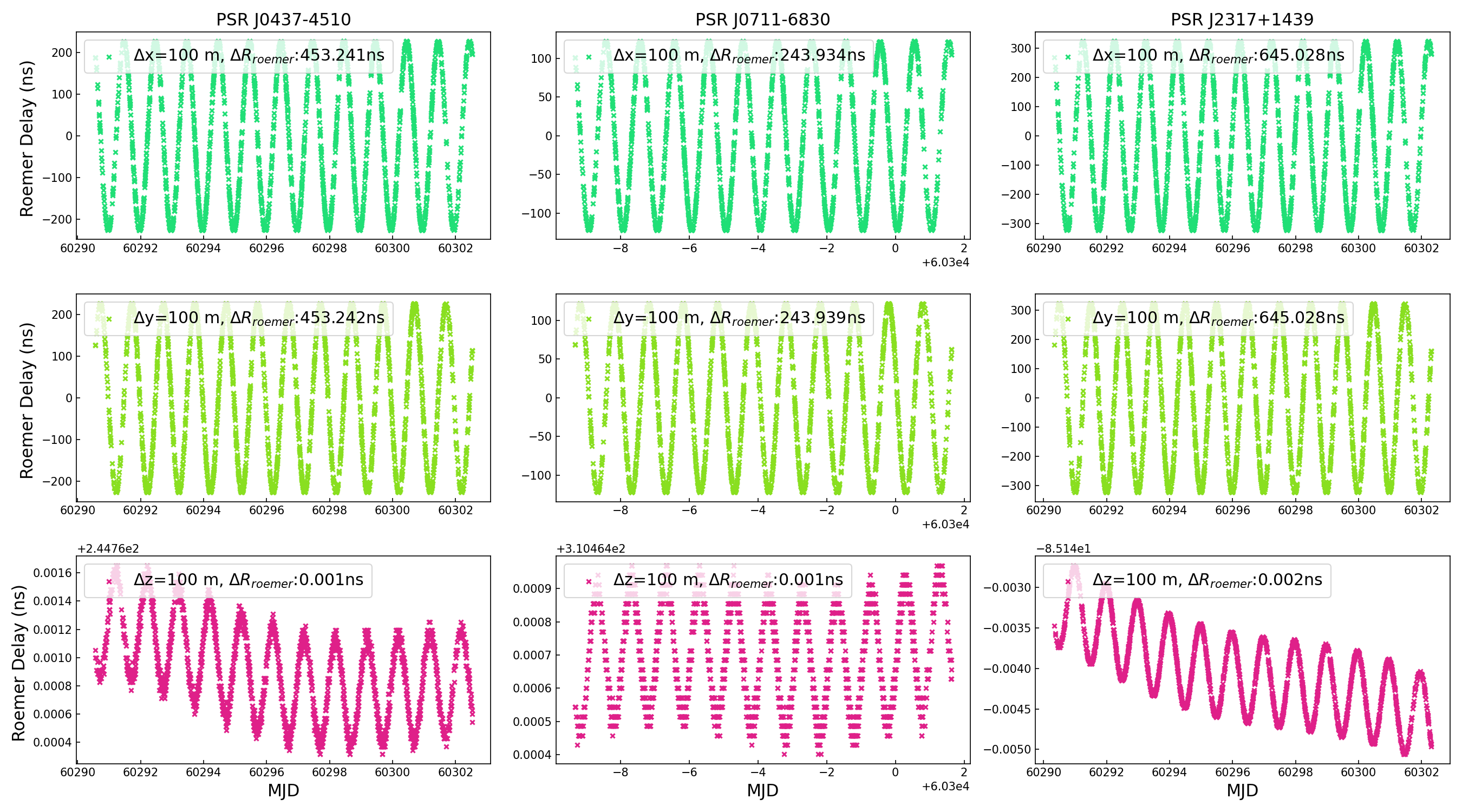}
\caption{The influence of station error on the calculation of 13-day Roemer delaywhen zenith angle changes.}
\label{fig4}
\end{minipage}
\end{figure}

Assuming that the station’s three-dimensional coordinates each have an error of 100 m and zenith angle varies within a certain range, Figure~\ref{fig4} illustrates the distribution of Roemer delay errors and their ranges for three pulsars over a 13-day period under three different scenarios. The results indicate that when the station errors are 100 m along the $x$-axis, $y$-axis, and $z$-axis directions, the induced delay errors and their ranges are identical for the $x$- and $y$-axis directions and are significantly larger than those along the $z$-axis direction. Furthermore, although the Roemer delay waveforms for different directions with the same error magnitude all exhibit similar quasi-periodicity (approximately one day), their initial phases differ.

\subsubsection{Shapiro Delay}

\begin{figure}[H]
\begin{minipage}[t]{0.999\linewidth}
\centering
\includegraphics[scale=0.20]{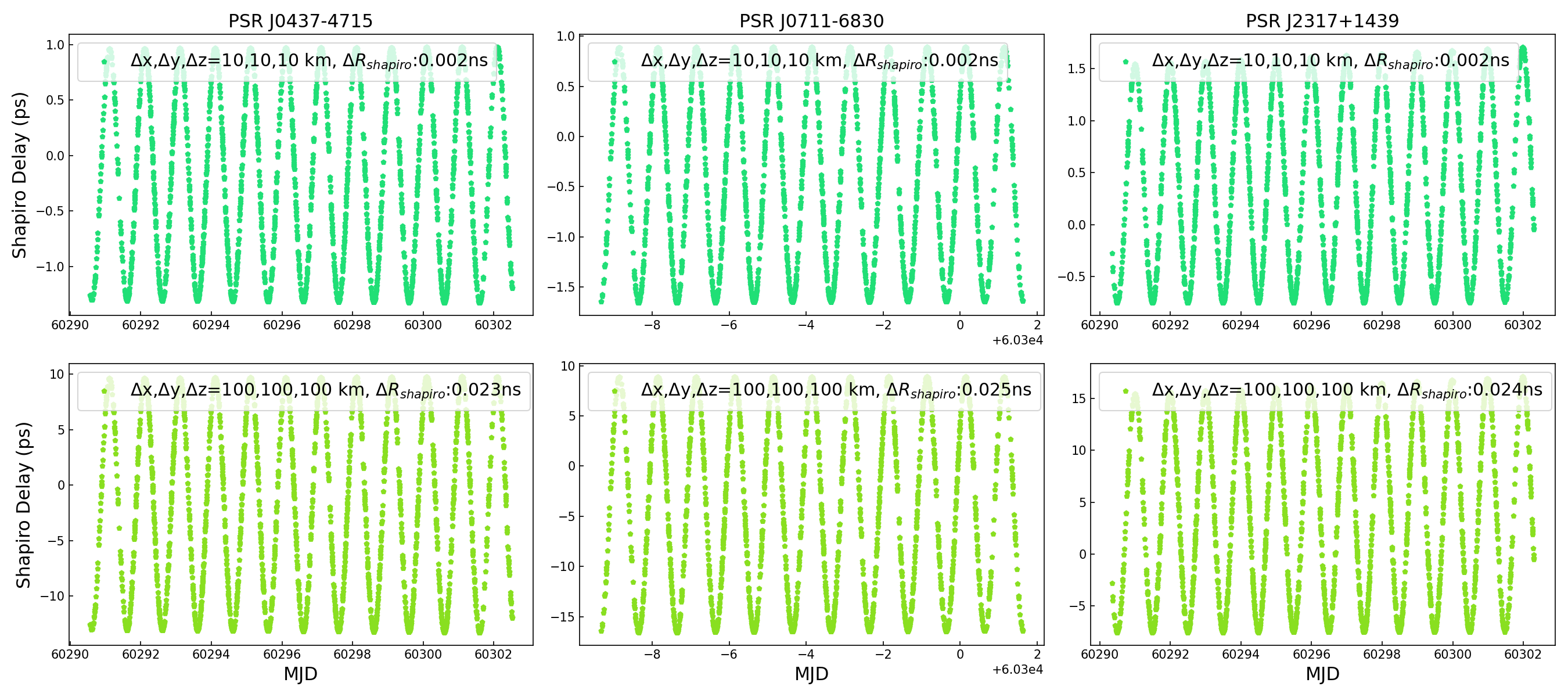}
\caption{The influence of station error on the calculation of 13-day Shapiro delay when zenith angle changes.}
\label{fig5}
\end{minipage}
\end{figure}

Assuming station three-dimensional coordinate errors at two levels, namely 10 km and 100 km, Figure~\ref{fig5} shows the distribution and range of Shapiro delay errors under these conditions when zenith angle changes. The results indicate that the Shapiro delay error within the solar system exhibits a linear relationship with the station position error. When the station coordinate error is 10 km, the Shapiro delay errors for all three pulsars are less than 2 ps. When the station position error increases to 100 km, the Shapiro delay errors for the three pulsars remain below 20 ps.

\subsubsection{Einstein Delay}

\begin{figure}[H]
\begin{minipage}[t]{0.999\linewidth}
\centering
\includegraphics[scale=0.20]{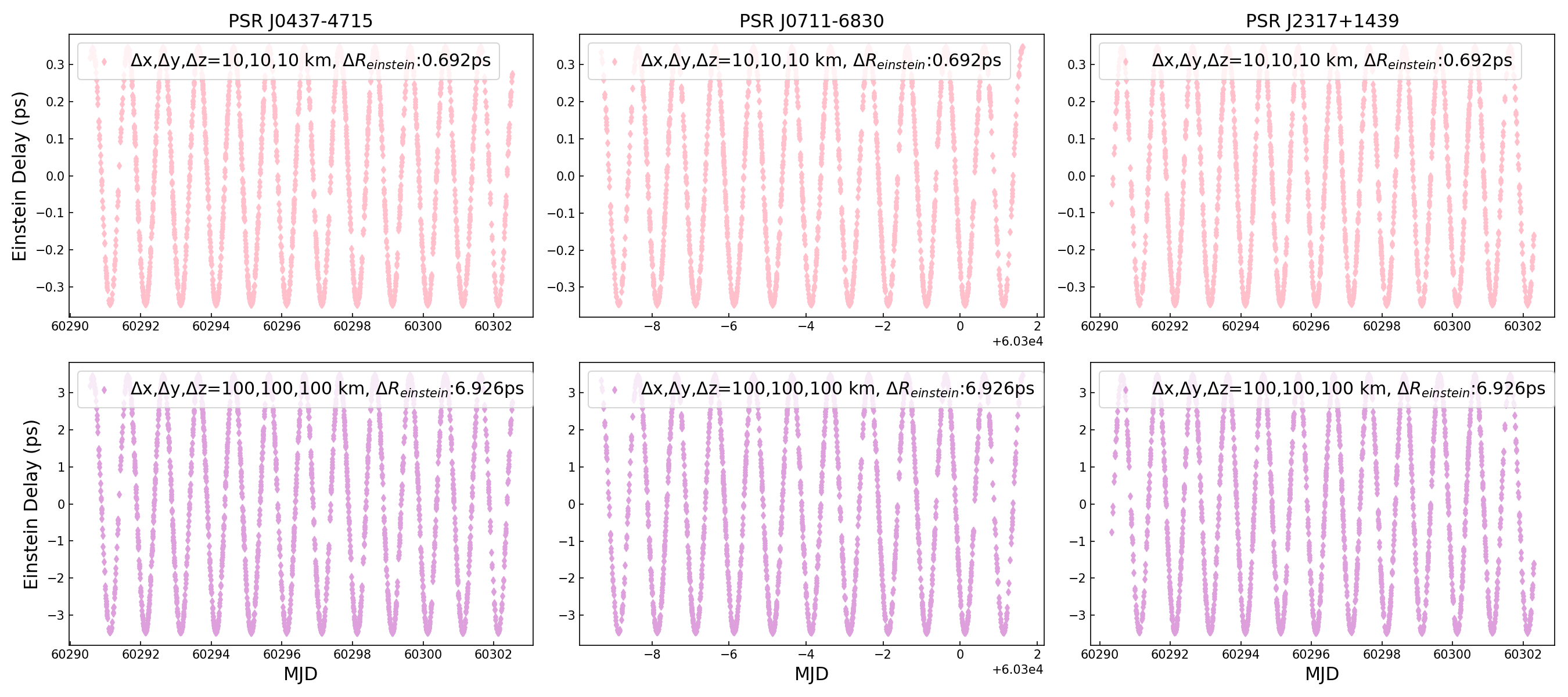}
\caption{The influence of station error on the calculation of 13-day Einstein delay when zenith angle changes.}
\label{fig6}
\end{minipage}
\end{figure}

In the case of zenith angle variation,figure~\ref{fig6} shows the magnitude and range of Einstein delay errors for the three pulsars when station three-dimensional coordinate errors of 10 km and 100 km are present, respectively. The results indicate that the Einstein delay error within the solar system exhibits a linear relationship with the station position error. When the station coordinate error is 10 km, the Einstein delay errors for all three pulsars remain below 0.5 ps. When the error increases to 100 km, the delay errors are on the order of a few picoseconds.

\subsubsection{Correction from TT to TCB} 
\begin{figure}[H]
\begin{minipage}[t]{0.999\linewidth}
\centering
\includegraphics[scale=0.20]{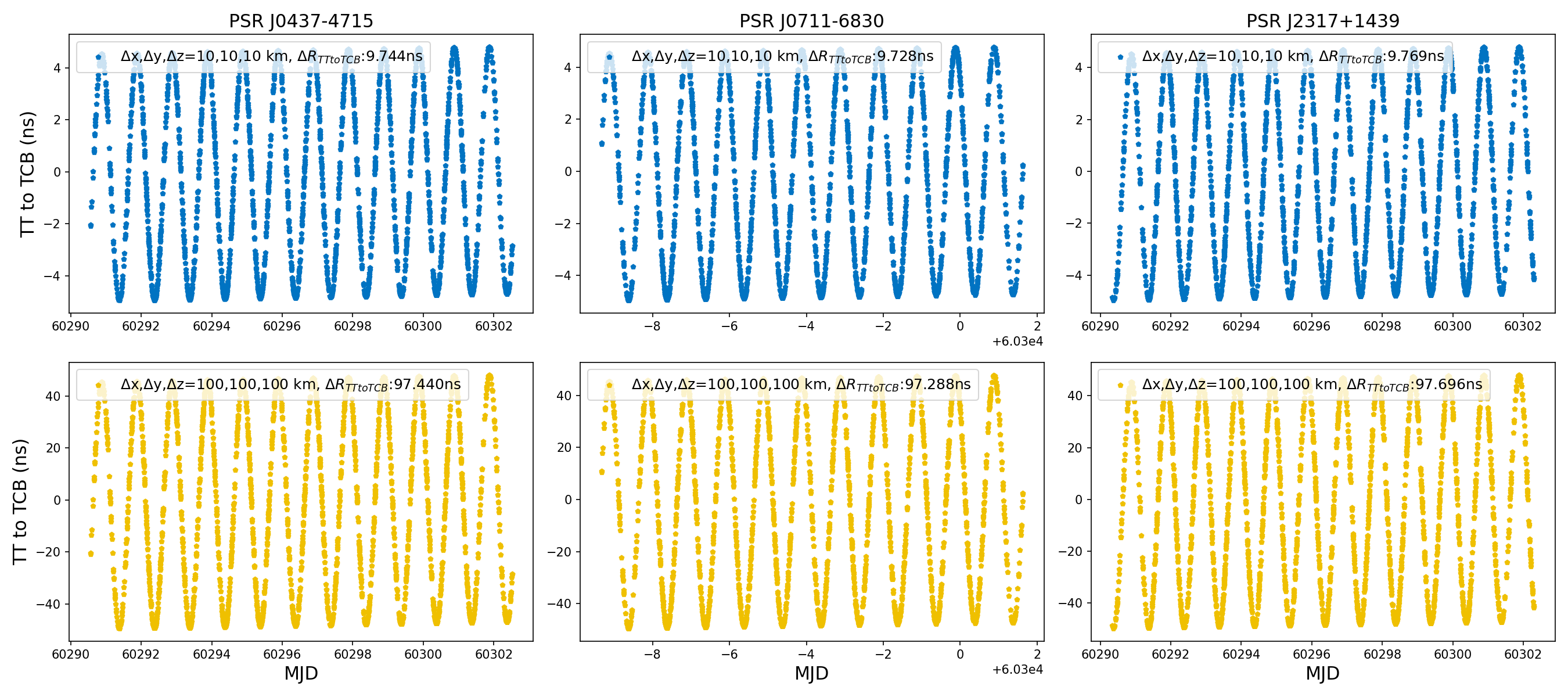}
\caption{The influence of station error on the calculation of 13-day  time delay caused by the transition from TT to TCB when zenith angle changes.}
\label{tt2tcb1}
\end{minipage}
\end{figure}

As illustrated in Figure~\ref{tt2tcb1}, when the zenith angle changes, the correction associated with the TT-to-TCB conversion exhibits a linear increase corresponding to the magnitude of the station coordinate errors. Specifically, for station coordinate errors of 10 km across all three spatial dimensions, the correction errors for the three pulsars are of similar magnitude, each remaining below 5 ns. Furthermore, when the station coordinate errors reach 100 km in all three dimensions, the correction errors for the three pulsars remain under 50 ns.Regardless of the magnitude of the station error, the error correction waveforms for the three pulsars exhibit a sinusoidal distribution, with initial phases that differ from one another. However, for any given pulsar, the initial phase remains the same.

\subsection{Analysis of the Simulation Results for 5-Year Continuous Sampling Data}

Table~\ref{Table3} summarizes the statistical results of the impact of station coordinate errors of different magnitudes on three millisecond pulsars during a five-year timing observation under the conditions of arbitrary change and constant zenith angle. The columns presented in the table are similar to those in Table~\ref{Table2}.Values without a superscript correspond to the case of a varying zenith angle, while values with a superscript \(z\) correspond to the case where the zenith angles of J0437$-$4715, J0711$-$6830, and J2317$+$1439 are assumed to remain constant at $14.21^{o}$,$35.56^{o}$,$47.79^{o}$,respectively.
The results indicate that as the station coordinate errors increase, the RMS, the range of Roemer delay error, the range of pre-fit residual error, and the range of ecorrection errors from TT
to TCB for all three pulsars exhibit a linear growth trend when
zenith angle changes. From the values with superscript \(z\) in each row of Table~\ref{Table3}, it is evident that, under a constant zenith angle, progressively changing the station coordinate error magnitude results in little to no variation in the RMS values, the Roemer delay, the residual range, and the TT-to-TCB conversion correction error. This indicates that, under a constant zenith angle, station coordinate errors have essentially no impact on pulsar timing. Detailed variations of each parameter will be discussed in subsequent sections.

\begin{table*} 
\center
\caption[]{The statistical analysis of the influence of station error on the five years pulsar timing of different pulsars.Values without a superscript refer to the case of arbitrarily varying zenith angle, whereas those with a superscript "z" indicate a constant zenith angle.}
\begin{tabular}{ccccccc}
\hline
 \multirow{2}{*}{Categories}& Psr Name &\multicolumn{5}{c}{($\Delta x$,$\Delta y$,$\Delta z$)(m,m,m)}\\
\cmidrule(lr){3-7}
   & (J2000) &(0,0,0) &(1,1,1) & (10,10,10)& (100,100,100)& (1000,1000,1000)\\
\hline
\multirow{3}{*}{RMS}&J0437$-$4715&0&0.002{/}0$^{z}$&0.022{/}0$^{z}$&0.224{/}0$^{z}$&2.245{/}0$^{z}$\\
\quad &J0711$-$6830&0 &0.001\text{/}$0^{z}$&0.012\text{/}$0^{z}$ &0.121\text{/}$0^{z}$&1.208\text{/}$0^{z}$ \\
 ($\mu$s)&J2317$+$1439&0 &0.003 {/}0$^{z}$&0.032{/}0$^{z}$&0.320{/}0$^{z}$&3.196{/}0.001$^{z}$\\
\multirow{3}{*}{$\Delta R_{roemer}$}&J0437$-$4715&0 &6.41{/}0$^{z}$&64.10 {/}0$^{z}$&640.96 {/}0.02$^{z}$&6409.56{/}0.19$^{z}$\\
 \quad &J0711$-$6830&0 &3.45\text{/}$0^{z}$&34.50\text{/}$0^{z}$&345.05\text{/}$0.03^{z}$&3450.45\text{/}$0.25^{z}$\\
 (ns) &J2317$+$1439&0 &9.12{/}0$^{z}$&91.24{/}0.03$^{z}$&912.37{/}0.32$^{z}$&9123.71{/}3.24$^{z}$\\
\multirow{3}{*}{$\Delta R_{pre-res}$}&J0437$-$4715&0 &6.76{/}0.31$^{z}$&64.46{/}0.31$^{z}$&641.25{/}0.65$^{z}$&6409.50{/}1.69$^{z}$\\
 \quad &J0711$-$6830&0 &3.68\text{/}$0.31^{z}$ &34.69\text{/}$0.31^{z}$ &345.32\text{/}$0.31^{z}$ &3451.09\text{/}$0.92^{z}$\\
 (ns) &J2317$+$1439&0 &9.52{/}0.32$^{z}$&91.49{/}0.32$^{z}$&912.94{/}0.67$^{z}$&9124.82{/}4.48$^{z}$\\
 \multirow{3}{*}{$\Delta R_{TT to TCB}$}& J0437$-$4715&0&0{/}0$^{z}$&0.01 {/}0$^{z}$&0.11{/}0.07$^{z}$ &1.12{/}0.69$^{z}$\\
 \quad &J0711$-$6830&0&0\text{/}$0^{z}$ &0.01\text{/}$0.01^{z}$ &0.11\text{/}$0.06^{z}$ &1.12\text{/}$0.60^{z}$\\
 (ns) &J2317$+$1439&0&0{/}0$^{z}$ &0.01{/}0.01$^{z}$&0.11{/}0.11$^{z}$ &1.12{/}1.05$^{z}$\\
\hline
\end{tabular}
\label{Table3}
\end{table*}


\subsubsection{TOA and Timing Residuals}

\begin{figure}[H]
\begin{minipage}[t]{0.999\linewidth}
\centering
\includegraphics[scale=0.20]{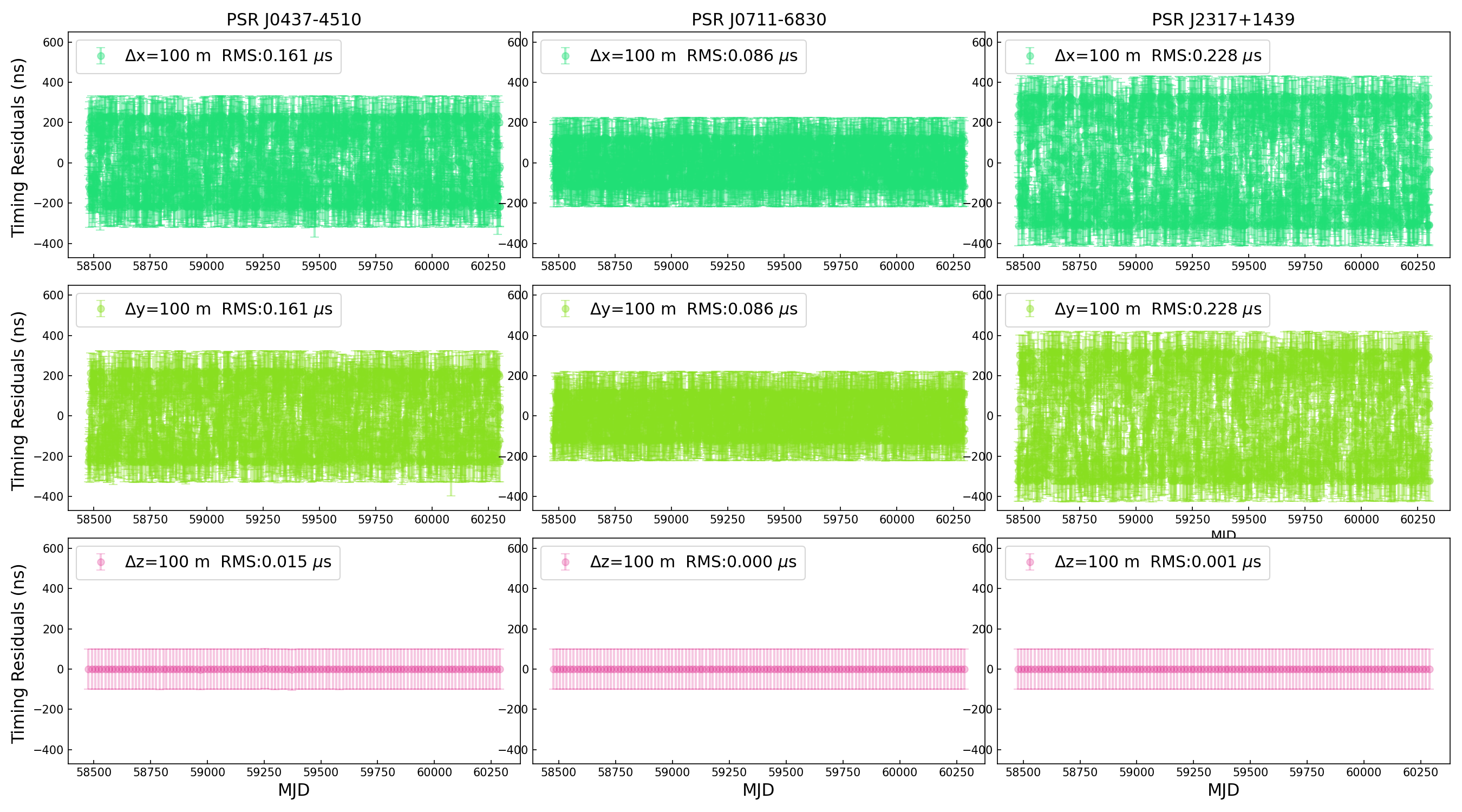}
\caption{The influence of station error on 5-year timing residual when zenith angle changes.}
\label{fig7}
\end{minipage}
\end{figure}

Under the condition that the zenith angle varies within a certain range and that the three-dimensional station coordinates each contain an error of 100 m, this paper analyzes the influence of station position errors on TOA and timing residuals of three pulsars over a continuous five-year span. The results are shown in Figure \ref{fig7}. Similar to the results in Figure \ref{fig3}, when the errors in the $x$-,$y$- and $z$- directions are all 100 m, the resulting changes in the RMS values are essentially identical in the $x$- and $y$-directions and are significantly greater than those in the $z$-direction. Different from Figure \ref{fig3}, the TOA data points caused by the three-dimensional station coordinate errors during the five-year period are distributed relatively uniformly; The residual values all cluster near zero and do not exhibit clear sinusoidal waveform characteristics.

\subsubsection{Roemer Delay}

\begin{figure}[H]
\begin{minipage}[t]{0.999\linewidth}
\centering
\includegraphics[scale=0.20]{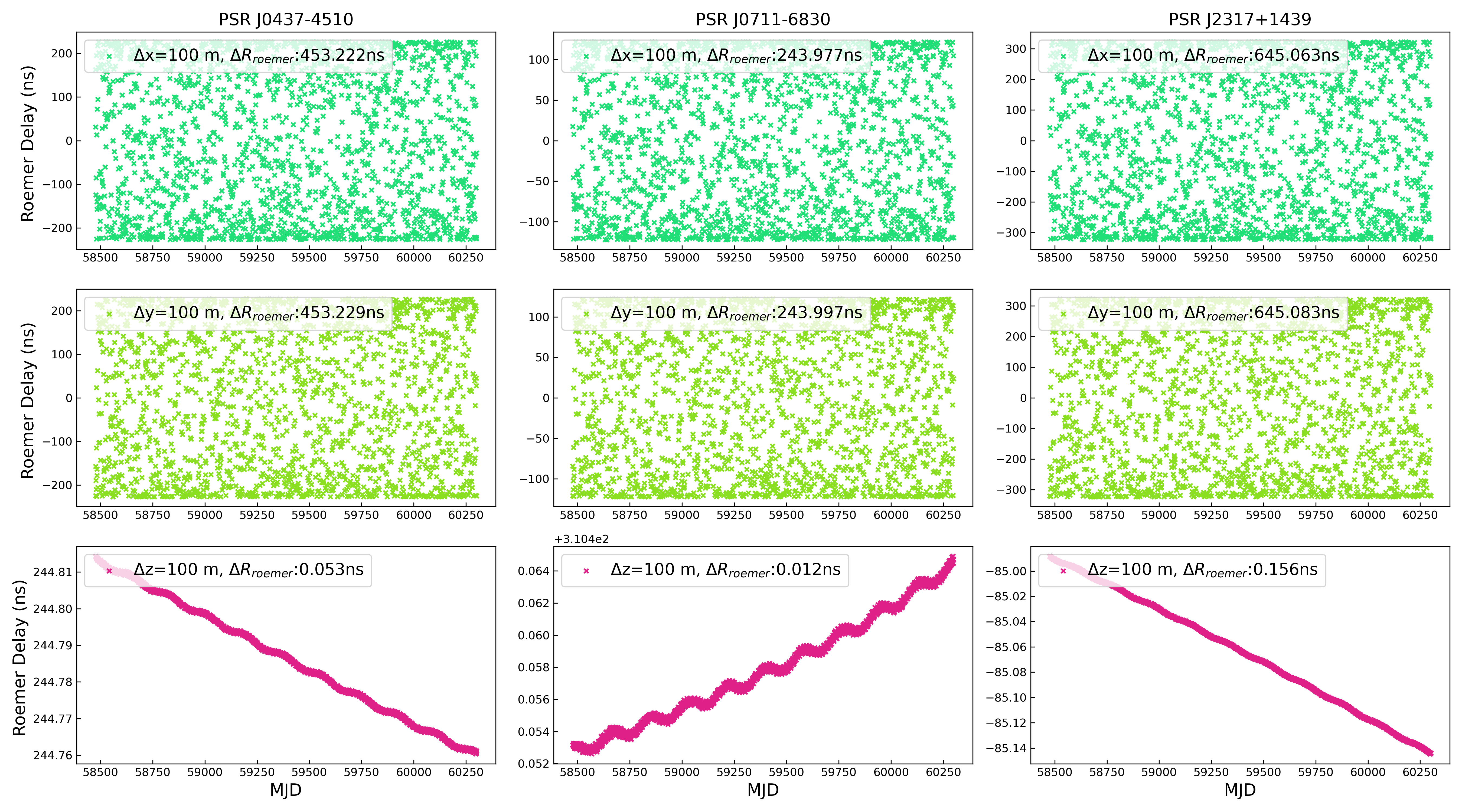}
\caption{The influence of station error on Roemer delay calculation for 5 years when zenith angle changes.}
\label{fig8}
\end{minipage}
\end{figure}

Assuming that zenith angle changes within a certain range and the coordinate error of the station is 100 meters,Figure~\ref{fig8} shows the distribution and range of Roemer delay errors for different pulsars over a five-year period in three scenarios. The results indicate that when the station errors are 100 m in the $x$-, $y$-, and $z$-axis directions, respectively, the delay errors and their ranges are identical for the $x$- and $y$- axis directions and are significantly larger than those for the $z$-axis. Unlike in Figure~\ref{fig4}, as the time span of station errors extends from 13 days to 5 years, the distribution of Roemer delay along the $x$- and $y$-axes no longer exhibits sinusoidal cosine waveforms but instead appears as a uniformly scattered distribution around zero. Meanwhile, the Roemer delay distribution along the $z$-axis shows sinusoidal/cosine waveforms with a linear trend. Fourier analysis further reveals that the Roemer delay waveforms of all three pulsars exhibit strict semi-annual and annual periodic oscillations.

\subsubsection{Shapiro Delay}

\begin{figure}[H]
\begin{minipage}[t]{0.999\linewidth}
\centering
\includegraphics[scale=0.20]{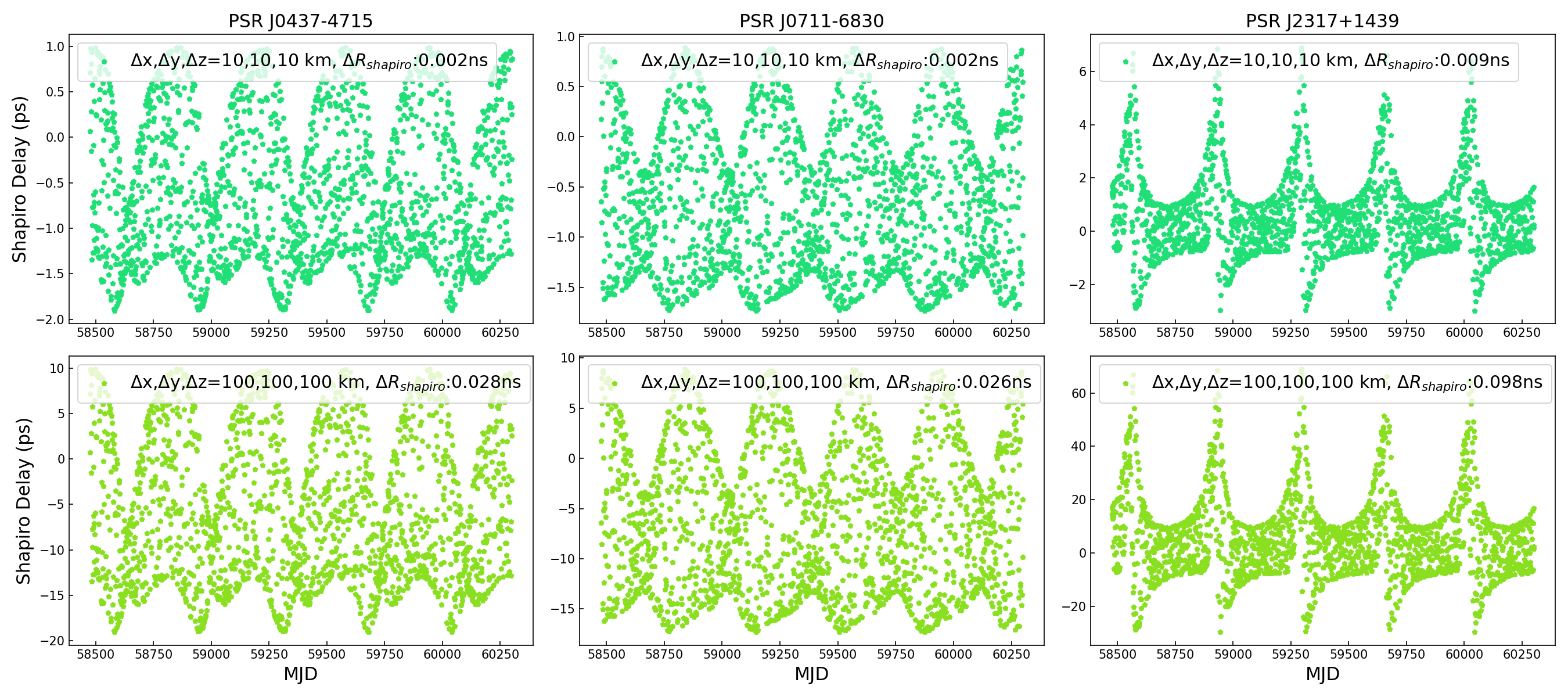}
\caption{The influence of station error on Shapiro delay calculation for 5 years when zenith angle changes.}
\label{fig9}
\end{minipage}
\end{figure}

Figure~\ref{fig9} illustrates the distribution and range of Shapiro delay errors for three pulsars over five consecutive years, considering station position errors of 10 km and 100 km when
zenith angle changes. The results indicate that the Shapiro delay error within the solar system scales linearly with the station position error. When the station coordinate error is 10 km, the Shapiro delay errors for all three pulsars are on the order of several picoseconds; when the error increases to 100 km, the delay errors remain only in the tens of picoseconds. Figure~\ref{fig9} also shows that, although the Shapiro delay distributions differ among the pulsars, they all exhibit a strict annual periodicity.

\subsubsection{Einstein Delay}

\begin{figure}[H]
\begin{minipage}[t]{0.999\linewidth}
\centering
\includegraphics[scale=0.20]{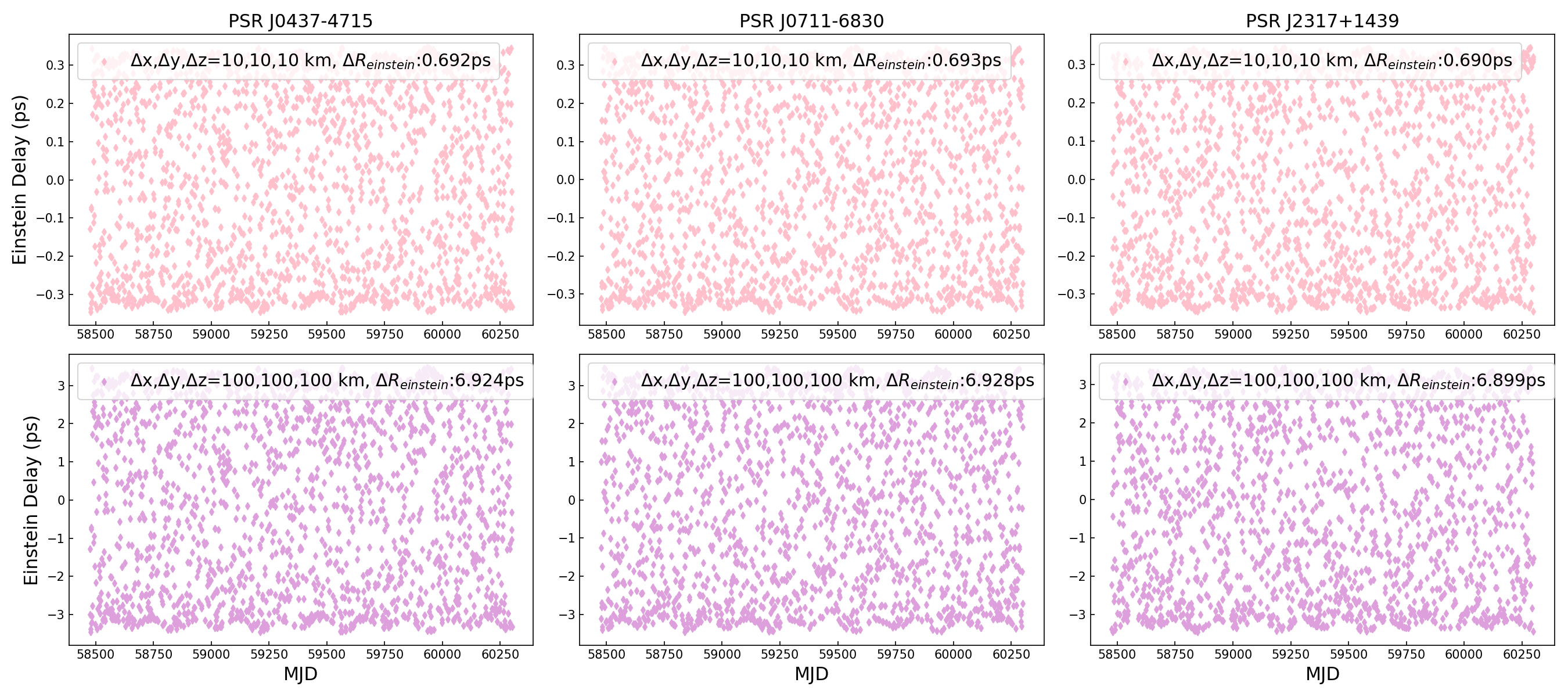}
\caption{The influence of station error on Einstein delay calculation for 5 years when zenith angle changes.}
\label{fig10}
\end{minipage}
\end{figure}

In the case of zenith
angle variation,Figure~\ref{fig10} presents the Einstein delay errors and their ranges for three pulsars under station position errors of 10 km and 100 km, respectively. The results demonstrate that the Einstein delay error within the solar system scales linearly with the station coordinate error. When the station position error is 10 km, the Einstein delay errors for all three pulsars remain below 1 ps. When the error increases to 100 km, the delay errors are still on the order of several picoseconds. Unlike the 13-day data shown in Figure~\ref{fig6}, the Einstein delay over the five-year period does not exhibit a distinct sinusoidal waveform; instead, it displays a uniformly scattered distribution around zero.

\subsubsection{Correction from TT to TCB} 
\begin{figure}[H]
\begin{minipage}[t]{0.999\linewidth}
\centering
\includegraphics[scale=0.20]{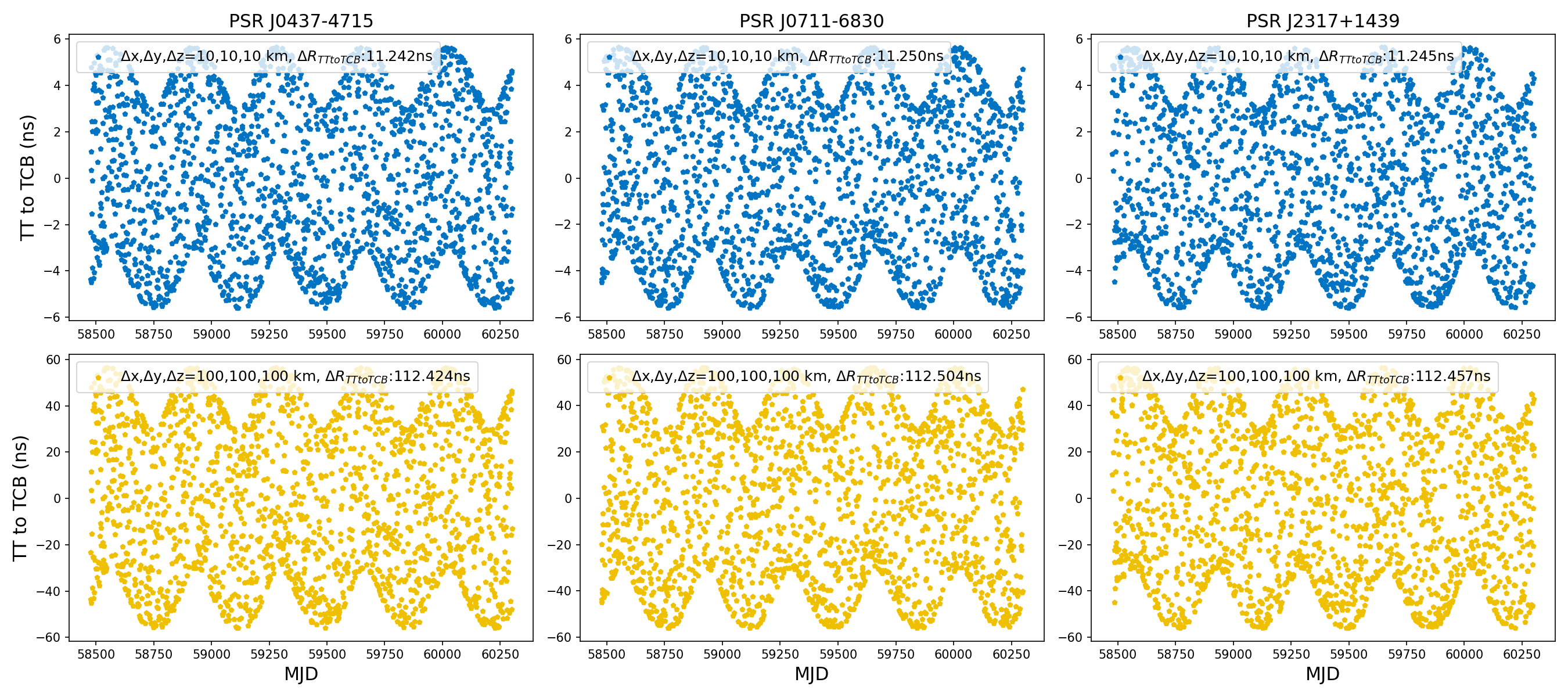}
\caption{The influence of station error on the calculation of 5-year time delay caused by the transition from TT to TCB when zenith angle changes.}
\label{tt2tcb2}
\end{minipage}
\end{figure}

Similar to Figure~\ref{tt2tcb1}, when the zenith angle changes, Figure~\ref{tt2tcb2} shows that the correction associated with the TT-to-TCB conversion increases linearly with the magnitude of the station coordinate error. When the station coordinate error is 10 km in all three spatial dimensions, the correction errors for the three pulsars are comparable in magnitude and remain below 6 ns. Furthermore, when the station coordinate error reaches 100 km in all three dimensions, the correction errors for the three pulsars remain below 60 ns. Unlike Figure~\ref{tt2tcb1}, regardless of the magnitude of the station error, the error waveforms for the three pulsars are irregular but nearly identical to one another, and all exhibit a strict annual periodicity.

\section{Discussion and Conclusion}

By employing the TEMPO2 software to simulate the magnitude of station coordinate errors, a comprehensive analysis of their impact on the precision of pulsar time-scale construction can be conducted. Comparing the data without the superscript z in rows 3, 4, and 5 of Tables \ref{Table2} and \ref{Table3}, it is evident that as the station coordinate errors increase proportionally, the RMS values of the three millisecond pulsars also exhibit an approximately proportional increase. This suggests that the stability of pulsar time decreases correspondingly. Under conditions of fixed station coordinate errors and the zenith angle varies, whether the observation duration is 13 days or 5 years, the variation trend of the RMS value for the same pulsar remains largely consistent. This indicates that once station position errors exist, their effect on the precision of pulsar time-scale construction persists over long timescales and is not mitigated by shorter observation periods or smaller datasets. Combined with the results shown in Figures~\ref{fig3} and ~\ref{fig7}, when the zenith angle changes and the single-axis errors in the three-dimensional station coordinates are numerically equal, the RMS values induced by errors in the $x$- and $y$-axes are essentially comparable to those caused by simultaneous errors of the same magnitude in the $x$-, $y$-, and $z$-axes, regardless of whether the observation duration is 13 days or 5 years. This demonstrates that the RMS contribution from $z$-axis coordinate errors alone is extremely small and negligible compared to the influence of the  $x$- and $y$-axes.A comprehensive analysis of rows 3 to 5 without a superscript in Tables 2 and 3, along with the computational results in Figures \ref{fig3} and \ref{fig7}, reveals through repeated observations and fittings that a linear relationship exists between station coordinate errors and the RMS variations of the three pulsars. When the observed zenith angle remains unchanged for a long time, that is, the values marked with letters in rows 3 to 5 of Table~\ref{Table3}, this relationship does not hold. This relationship can be expressed as: $RMS \approx \sqrt{(a\Delta x)^2 + (b\Delta y)^2 + (c\Delta z)^2}$, where the parameters $a$, $b$, and $c$ are the linear coefficients corresponding to the three spatial directions, and the units of coordinate error and RMS are meters (m) and seconds (s), respectively. Table~\ref{Table4} lists the values of these linear function parameters for the three pulsars during the 13-day and 5-year observation periods, with all parameters rounded appropriately. The data in the table show that the values of parameters $a$ and $b$ are generally similar, while parameter $c$ is one or two orders of magnitude smaller than $a$ and $b$ and is positive in all cases. This confirms that station errors in all three spatial directions contribute to the RMS, but the contribution from the $z$-axis direction is relatively minor. Based on the functional relationship between the residuals of PSR J0437$–$4715 and the three-dimensional coordinate errors, it can be inferred that, when zenith angle
changes, in order to keep the increment in residuals caused by station errors below 1 ns, the three‑dimensional coordinate errors of the Parkes telescope must be maintained within a range of 0.62 to 5.85 meters.
\begin{table*}
\center
\caption[]{The parameter fitting results of the linear relationship between station error and RMS of three millisecond pulsars.}
\begin{tabular}{ccccccc}
\hline
 Pulsar name &\multicolumn{3}{c}{13-day data fitting}&\multicolumn{3}{c}{5-year data fitting}\\ 
\cmidrule(lr){2-4}
\cmidrule(lr){5-7}
 (J2000) & $a$ & $b$ & $c$ & $a$ & $b$ & $c$ \\
\hline
 J0437$-$4715& $1.61\times{10}^{-9}$  &$1.57\times{10}^{-9}$  &$1.71\times{10}^{-10}$ &$1.59\times{10}^{-9}$  &$1.59\times{10}^{-9}$  &$1.54\times{10}^{-10}$\\
 J0711$-$6830& $8.59\times{10}^{-10}$ &$8.59\times{10}^{-10}$ &$2.07\times{10}^{-11}$ &$8.56\times{10}^{-10}$ &$8.56\times{10}^{-10}$ &$2.02\times{10}^{-11}$\\
 J2317$+$1439& $2.28\times{10}^{-9}$  &$2.28\times{10}^{-9}$  &$1.36\times{10}^{-11}$ &$2.27\times{10}^{-9}$  &$2.27\times{10}^{-9}$  &$2.06\times{10}^{-11}$\\
\hline
\end{tabular}
\label{Table4}
\end{table*}

Based on the analysis of rows 6 to 8 without the superscript z in Tables \ref{Table2} and \ref{Table3}, it is observed that as the station coordinate errors increase proportionally, the range of Roemer delay errors for different pulsars also shows a corresponding multiplicative growth. This trend remains unaffected by the duration of observation. Combining with Figures~\ref{fig4} and ~\ref{fig8}, it is found that the range of Roemer delay errors caused by single-axis errors ($x$ or $y$) of the station reference point in the Earth-fixed coordinate system remains largely consistent across both the 13-day and 5-year timescales when zenith angle varies. However, for single-axis errors in the z-direction, the Roemer delay error over the 5-year period is significantly larger than that over the 13-day period. This indicates that the influence of station coordinate errors on the Roemer delay includes not only a component with a period of one day but also a component with a period of one year, with the latter being much larger than the former—fully consistent with theoretical model predictions. Due to the non-uniformity of the apparent annual motion in true solar time, while mean solar time (commonly used in daily applications) moves more uniformly along the equator, the equation-of-time curve exhibits both semi-annual and annual quasi‑periodic characteristics. This explains why the Roemer delay pattern resulting from $z$-axis errors over 5 years displays both semi‑annual and annual periodicities.

According to the data in rows 9 to 11 without superscript letter z of Tables \ref{Table2} and \ref{Table3}, as station coordinate errors increase multiplicatively, the range of pre-fit residuals for different pulsars shows a linear growth trend, regardless of whether the error duration is 13 days or 5 years. A further vertical comparison between rows 6–8 and rows 9-11 in the tables reveals that the difference between the range of pre-fit residuals and the corresponding range of Roemer delay errors is minimal or even equal,when the zenith angle varies. Kendall correlation analysis performed on these two sets of data yields six correlation coefficients $r$ all equal to $1.67\%$, with corresponding $p$-values of $100\%$, indicating a significant positive correlation between the range of residuals induced by station coordinate errors and the range of Roemer delay errors. This result further demonstrates that, under current pulsar timing precision and when the zenith angle changes, station coordinate errors primarily affect the pulse arrival time by influencing the Roemer delay term, thereby contaminating the pulsar timing residuals. This conclusion is fully consistent with Equations \ref{eq1}-\ref{eq3}.

Comparing the last three rows of Tables \ref{Table2} and \ref{Table3}, it is observed that when the zenith angle changes, regardless of whether the observation duration is 13 days or five years, the correction error for the TT-to-TCB conversion increases slowly and linearly with increasing station coordinate error. The magnitude of the correction error over the five-year period is slightly larger than that over the 13-day period for the three pulsars, and the correction errors for the three pulsar are exactly the same for the same period.Under a constant zenith angle, the five‑year TT‑to‑TCB conversion correction error due to station coordinate errors for the two pulsars (excluding J2317$+$1439) is smaller than the corresponding 13‑day error with a varying zenith angle. This error, however, seems to increase linearly. These findings suggest that station coordinate errors affect the coordinate conversion correction irrespective of zenith angle variations. In general, such errors are small and can be considered negligible under some circumstances.

Due to the limitation of the telescope's elevation angle, the zenith angle during actual observations must be smaller than the telescope's maximum zenith angle. At observing sites where the zenith angle varies only within a very limited range, such as Arecibo or FAST, the effect of station coordinate errors on pulsar timing is further reduced. Currently, even for IPTA projects and datasets with the highest timing precision requirements, the effect of zenith angle variations on pulsar timing has not been taken into account. This means that the conclusions in the text are likely applicable to the vast majority of observatories, except for FAST.

This study systematically elucidates the quantitative influence of station coordinate errors on the precision of pulsar time-scale construction. It clearly identifies errors along the $x$- and $y$-axis directions within the Earth-fixed coordinate system as the primary factors, with their impacts being largely comparable. In contrast, errors along the $z$-axis have a relatively minor influence, consistent with theoretical expectations.Under long‑term zenith‑angle variations, station coordinate errors primarily affect the residuals of pulsar signal arrival times at the SSB through the Roemer delay, while the effects of other correction and delay terms, such as the Shapiro and Einstein delays, are negligible. Under current equipment precision and engineering test conditions, station coordinate errors range from millimeters to kilometers. Within this range, the linear relationship between station coordinate errors and both RMS and Roemer delay errors indicates that the mechanisms and magnitudes of their impact on pulsar time-scale construction precision have been thoroughly characterized.

Given the strong agreement between the theoretical model and experimental results, future research could explore using various error components in pulsar time-scale construction precision to inversely estimate the range of station coordinate errors. This approach holds significant scientific value and has important practical implications for both military and civilian applications. In routine pulsar time research and engineering experiments, it is crucial to prioritize the regular, precise measurement of station positions and the systematic analysis of related errors. Proactive efforts in error calibration and mitigation are essential to fully realize the potential of constructing high-precision pulsar time.

The execution of this study is both necessary and innovative. First, as a core technology for future deep-space navigation and autonomous time referencing, the precision of pulsar timing directly determines the feasibility of related applications. Station coordinate errors, a critical yet insufficiently analyzed source of error, urgently require thorough investigation and calibration. Second, this research departs from the conventional approach of treating station coordinates as perfectly known quantities. Instead, it systematically quantifies them as key variables, thereby addressing a gap in the error system that affects the precision of pulsar time-scale construction. Third, it establishes a comprehensive transmission chain from coordinate errors to delays, corrections, residuals, and ultimately time stability. This clearly elucidates the mechanisms and magnitudes of error influence, laying the groundwork for model-guided precise calibration in engineering applications. Finally, it proposes a novel reverse-application concept: using high-precision pulsar timing to infer the stability of the terrestrial reference frame. This approach fosters deep interdisciplinary integration between astronomy and geodesy, thereby broadening the application scope of pulsar time.

In summary, this study not only offers a direct technical approach to enhancing the precision of pulsar time-scale construction but also provides novel insights through a unique perspective and methodology. These contributions hold significant value and potential applications in related fields. Future research may focus on translating these findings into practical pulsar time calibration algorithms and engineering standards, thereby facilitating the development and implementation of next-generation spatiotemporal reference systems.

\section*{acknowledgements}
We express our gratitude to the other members of the project team for their for their
hard work, dedication, support, and assistance. 
This work was supported by the Natural Science Basic Research Program of Shaanxi (2024JC-YBQN-0036), the National Natural Science Foundation of China (11973046), the Project of the "West Light" Talent Training Program (XAB2021YN27), the Strategic Priority Research Program (Category A) of CAS (XDA0350502), and the National SKA Program of China (2020SKA0120200).

\section*{AI disclosure statement}
Deepseek was utilized to perform language and grammar assessments within the article. The authors meticulously reviewed, edited, and modified the texts produced by Deepseek according to their own standards, thereby assuming full responsibility for the final content of the publication.

\section*{Author Contributions}
Yuping Gao led the project. Chengshi Zhao conceived and developed the original idea. Zurong Zhou designed and conducted the study, and drafted the manuscript. Jianping Yuan and Wei Han contributed to the conceptual refinement and performed the experiments with assistance from Yue Hu and Shijun Dang. Shougang Zhang, Na Wang, Jingbo Wang, Minglei Tong, and De Wu reviewed and edited the paper. All authors read and approved the final version of the manuscript.

\section*{Declaration of Interests}
The authors declare no competing interests.






\end{document}